A first principles study of high pressure phase diagram of bismuth at 0 ~14 GPa based on structure search and quasi-harmonic approximation

Ya-Fan Zhao*[a,b], Shou-Rui Li[c], Xing-Yu Gao[a,b], Ming-Feng Tian[a], Hai-Feng Song*[a,b]

a) Institute of Applied Physics and Computational Mathematics, Beijing 100088, China

b) Software Center for High Performance Numerical Simulation, China Academy of Engineering Physics, Beijing 100088, China

c) Institute of Fluid Physics, China Academy of Engineering Physics, Mianyang 621900, China

**Abstract**

We have performed first principles study for bismuth crystal structure at pressure from 0~14 GPa based on structure search and quasi-harmonic approximation. A new stable phase with *Pbcm* symmetry is predicted by structure search method. We find that the most stable structures of bismuth at 0 K are $R\bar{3}m$(0 ~ 3.29 GPa), *Pbcm* (3.29 GPa ~ 4.91 GPa), *Cmce* (4.91 GPa ~ 10.57 GPa) and $Im\bar{3}m$ ( > 10.57 GPa), respectively. By quasi-harmonic approximation, we predicted the phase diagram of bismuth from first principles calculations. We found that the phase transition pressure reduces with increasing temperature. Our calculation agrees with the trend of experimental phase diagram. The *P4/ncc* model structure for the incommensurate Bi-III phase is not a stable phase in our calculation. A better model for the Bi-III phase is still needed. We also note that the spin-orbital interaction is very important for phase-diagram simulation of bismuth. By using first principle based structure search method, we successfully determine the low temperature high pressure phase diagram of bismuth, showing that the structure search method can effectively find the most stable structure of given material at high pressure even with high Z elements.

**1. Introduction**

For decades, the phase transition of bismuth at high pressure has drawn extensive attention, and is widely used as a calibration standard for high-pressure experiments with multi-anvil apparatus (MAA). Under ambient temperature, the structure of bismuth is proposed to transform from Bi-I phase ($R\bar{3}m$ symmetry) to Bi-II phase at 2.54 GPa.[1] Bi-II phase is stable in a narrow pressure range from 2.54 GPa to 2.7 GPa. When the pressure increases, Bi-II phase transforms to Bi-III phase. The Bi-III phase was previously reported to be of P4/n symmetry with 10 atoms in a unit cell.[2] Recent experiments have confirmed that Bi-III possesses an incommensurate structure with a body-centered-tetragonal (bct) "host" and a bct "guest" component[3] . In addition, it was reported that another two phase transitions exist between Bi-III to Bi-III' at pressure 4.3 GPa, and further to Bi-III" at 5.3 GPa with resistance measurements.[4] However, recent diffraction experiments by Chen et al have not observed these two phase transitions.[5] The Bi-III structure transforms to bcc structure at 7.7 GPa.[6] The bcc structure is proposed to be stable at higher pressure until 220 GPa. [6-9]

Apart from these phases at room temperature, several stable high temperature phases were also reported. In the 1990s, Chen and Iwasaki et al reported a Bi-IV structure, and suggested a monoclinic *P2₁/c* unit cell containing 8 atoms with lattice constants a = 6.468(5) Å, b = 6.578(5) Å, c = 6.468(5) Å and β = 118.88(6)° at 3.9 GPa and 503 K. [9-10] A few year later, V. F. Degtyareva reported that the Bi-IV phase had a Cmce structure at 3.9 GPa and 503K. [11] The *Cmce* phase was also confirmed by Chaimayo et al at 3.2GPa and 465K in 2012. [12]

At low temperature, the phase transition sequence is different from the high temperature phase transition. Previous studies by Homan et al suggested that the Bi-II phase, which is stable at room temperature under pressure 2.5~2.8 GPa, is not stable when the temperature is lower than 160 K. [13] In the phase diagram of bismuth, there are about 10 possible solid phases. But after so many years of study, only the crystal structures of phases Bi-I, Bi-II, Bi-III, Bi-IV and Bi-V are clearly known so far. The existence and the crystal structures of other possible phases are still unknown. The phase transitions of different bismuth phases at low temperature remain unclear.

Above all, it is necessary to clarify the disagreement and make further investigation on the phase diagram of bismuth. In recent years, structure search methods based on first principle calculation have been well developed [14-16] and widely used, including crystal structures at high pressure [17-19]. Structure search of bismuth based on first principles calculations were also performed using the CALYPSO method[20] and USPEX [21]. Here we also perform structure prediction based on first principle calculations. The article is organized as below. In the Method section the computational details are introduced. Following section is Discussion of the calculation results. We end the article with Conclusions.

## 2. Method

The structure prediction is performed using the CESSP program[22-25]. The structures search is performed using Basin hopping algorithm[26], which has been shown to be efficient in cluster[27-28] and crystal structure[29] search. The number of atoms in one unit cell could be 2, 3, 4, 5, 6, 8 and10. Structure search calculations with lager unit cells are beyond our computing capacity. In order to get the most stable crystal structure at different pressure, the structure search was performed under a series of pressure. According to the phase diagram of bismuth at low temperature, the selected pressure points are 0 GPa, 3 GPa, 4 GPa, 5 GPa, 6 GPa, 8 GPa and 13 GPa.

For the geometry optimization, we used VASP code[30]. To compare the energy of structures, we used the GGA-PBE functional[31-32] and the PAW method[33]. A plane wave cut off energy of 500 eV and KSPACING of $0.02*2\pi$ Å$^{-1}$ was used so as to get an energy convergence of less than 1 meV. The number of valence electrons considered in the calculation is 15 for each Bi atom. The SCF convergence is set to be 1.0E-8. The conjugate gradient (CG) method is employed during the geometry optimization and the force convergence criteria is 0.0005 eV/ Å. Smearing method of Methfessel-Paxton order 1 with SIGMA=0.2 is used during the geometry optimization. For total energy calculation, tetrahedron method with Blöchl corrections smearing method is applied. Previous work suggested that the spin-orbital interaction (SOI) plays an important role in the accurate energy of bismuth [34]. However, the inclusion of SOI greatly increases the computational cost, so the spin-orbital coupling effect is only considered in the static energy calculation. If the most stable structure changed with increasing pressure, phase transition is expected to be observed, the transition pressure is carefully located using bisection method. At the transition pressure, the two different phases should have equal formation enthalpies. The phonopy program (引用 phonopy) is applied to get the phonon spectra of different structures. The quasi-harmonic approximation method was employed to simulate the free energy of different structures at different temperatures.

The DFT calculation only considered the enthalpy of the structure. In order to compare our calculation with the experimental phase diagram at limited temperature, we performed phonon calculation of the structures and used the quasi harmonic approach to simulate the properties of Bi crystal at finite temperature.

## 3. Results and Discussion

The most stable structures of bismuth at different pressures are determined. Here we summarize the results from the structure search.

### 3.1 Structure of Bi at different pressures at 0 K

#### 3.1.1 Crystal structures of Bi at 0.0 GPa

At 0 GPa, the most stable structure is of $R\bar{3}m$ symmetry (Space Group # 166). The lattice constants from calculation are $a = b = 4.579$ Å, $c = 12.184$ Å, $\gamma = 120°$, $Z = 6$. The calculated parameters are in good agreement with the experimental value of $a = b = c = 4.7459$ Å, $\alpha = \beta = \gamma = 57.237°$. The density of the $R\bar{3}m$ structure obtained from first principle is 9.413 g/cm$^3$. The lattice constants and energy of each structure are listed in Table 1.

The second most stable structure is of *Imma* symmetry (*Imma*-S1) (Space Group #74), with lattice constants of $a = 6.522$ Å, $b = 4.558$ Å, $c = 4.955$ Å. The density of this structure is 9.469 g/cm$^3$. The energy of the structure is only 1.39 meV/atom higher than the $R\bar{3}m$ structure. The Wyckoff positions and coordinates of the new structures are listed in Table 2.

The third most stable structure is of *I4/mmm* symmetry (Space Group #139). The lattice constants are $a = b = 4.609$ Å, $c = 6.902$ Å, with a calculated density of 9.582 g/cm$^3$. Its energy is 5.8 meV/atom higher than the $R\bar{3}m$ structure.

A *Pbcm* structure is the fourth most stable structure with $a = 6.354$ Å, $b = 6.738$ Å, $c = 6.354$ Å, with a calculated density of 10.21 g/cm$^3$. Its energy is 48.3 meV/atom higher than the most stable $R\bar{3}m$ structure.

The *C2/m* structure, which is reported to be stable at 2.54GPa~2.7GPa, is also built manually and optimized. However, the optimized structure is another *Imma* structure with $a = 6.384$ Å, $b = 6.327$ Å, $c = 3.368$ Å (*Imma*-S2). The density of the *Imma* structure is 10.206 g/cm$^3$. The energy of this structure is 50.9 meV/atom higher than the $R\bar{3}m$ structure. The result agrees with previous DFT work by Häussermann [35] and experiment that Bi-I to Bi-II transition disappears below 200K.

At 0 K, the most stable bismuth structure is $R\bar{3}m$ structure, which agrees with the experiment.

#### 3.1.2 Crystal structures of Bi at 3.0 GPa

At 3.0 GPa, the $R\bar{3}m$ structure is still the most stable one. The lattice constants are $a = b = 4.514$ Å, $c = 11.534$ Å. The density of the $R\bar{3}m$ structure increases to 10.231 g/cm$^3$.

The *Imma* (S1) structure is still the second most stable one. The lattice constants are compressed to $a = 4.505$ Å, $b = 6.416$ Å and $c = 4.689$ Å. The density of this structure is 10.244 g/cm$^3$. Its energy is 1.7meV/atom higher than the $R\bar{3}m$ structure.

The *Pbcm* structure becomes the third most stable. The lattice constants at 3 GPa is $a = 6.2811$ Å, $b = 6.6180$ Å, $c = 6.1951$ Å. The density of *Pbcm* structure increases to 10.781 g/cm$^3$. The stability of the *Pbcm* structure increases as the pressure increases, and its energy is only 3.7meV/atom higher than the *R$\bar{3}$m* structure.

As a comparison, the high temperature Bi-IV phase with *Cmce* symmetry is constructed and optimized. The energy of this structure is 21.5 meV/atom higher than the *R$\bar{3}$m* structure. The Bi-III phase was modeled with a *P4/ncc* structure with 32 atoms, which is also adopted by previous DFT work[21, 35]. The enthalpy of the *P4/ncc* structure is 22.9 meV/atom higher than the *R$\bar{3}$m* structure.

At 3 GPa, the *R$\bar{3}$m* structure is still most stable. However, the energy difference of the *Pbcm* structure and the *R$\bar{3}$m* structure is greatly reduced at this pressure.

**3.1.3 Crystal structures of Bi at 4.0 GPa**

At 4.0GPa, a phase transition is observed. The *Pbcm* structure becomes the most stable structure. The density of the *Pbcm* structure increases to 10.994 g/cm$^3$. At this pressure, the enthalpy of the *R$\bar{3}$m* structure is 8.2 meV/atom higher than the *Pbcm* structure. The density of the *R$\bar{3}$m* structure at this pressure is 10.445 g/cm$^3$. There is a density increase of 5.3% by phase transition from *R$\bar{3}$m* to *Pbcm*, which is quiet close to previous experiment value of Yoneda and Endo[36], who reported a volume change of 4.9% for Bi-I to Bi-II transition. Using the interpolation method, the *R$\bar{3}$m* to Pbcm phase transition pressure at 0 K is determined to be 3.29 GPa. This is in disagreement with previous calculation by Häussermann,[35] whose research indicated that the *R$\bar{3}$m* Bi-I transfers to Bi-III directly at 3.54 GPa. It was reported by Compy *et al* that at low temperature[37], the Bi-II phase is not stable and there might exist a new-phase II of bismuth. We suggested that the *Pbcm* structure might be a new phase II of bismuth at low temperature.

The second most stable structure is a *C2/c* structure. Its energy is 0.6 meV/atom higher than the *Pbcm* structure. The relative stability of the *Cmce* structure and the *P4/ncc* structure also increases, and their enthalpies are 8.4 and 10.6 meV/atom than the Pbcm structure respectively. Lattice constants and energies of other meta-stable structures are listed in Table 1.

At 4 GPa, the most stable structure is the *Pbcm* structure. The *R$\bar{3}$m* → *Pbcm* phase transition pressure is 3.29 GPa at 0 K. The *P4/ncc* structure and the *Cmce* structure are less stable.

**3.1.4 Crystal structures of Bi at 5.0 ~10.0 GPa**

At 5.0 GPa, another phase transition is observed. The phase transition is from the *Pbcm* structure to the Bi-IV *Cmce* structure, which has been reported by Degtyareva as a high-pressure high temperature phase Bi-IV.[11]. The density of the most stable *Cmce* structure is 11.659 g/cm$^3$, while the density of the Pbcm structure is 11.186 g/cm$^3$. The density is increased by 4.2% because of the phase transition. The Bi-III modeled by the *P4/ncc* structure is less stable than the *Cmce* structure by 3.1 meV/atom . The density of the *P4/ncc* structure is 11.619 g/cm$^3$. The enthalpy of the *R$\bar{3}$m* structure is now 19.9 meV/atom higher than the *Cmce* structure. The *Pbcm* to *Cmce* phase transition pressure is determined to be 4.91 GPa at 0 K.

Previous work by Chen et al [10] reported a *P2$_1$/c* structure Bi-IV phase. As a comparison, we also build a *P2$_1$/c* structure from Chen's work and optimized the structure with DFT method. Surprisingly, we found that the *P2$_1$/c* structure is actually unstable and is automatically transformed to the *Cmce* structure after geometry optimization. If first principles calculation was performed when Chen *et al* found the *P2$_1$/c* structure, the *Cmce* structure would be found much earlier.

Though the *Cmce* structure is supposed to be stable at high temperature, we found that it is the most stable structure at 0 K and pressure range from 5.0 GPa to 10.0 GPa. At 10.0 GPa, the *Im$\bar{3}$m* structure, which is supposed to be the most stable structure from 7.7 GPa, is still 1.1 meV/atom higher in energy than the *Cmce* structure. The enthalpy of the *P4/ncc* model structure, however, is 6.3 meV/atom higher than the *Cmce* structure at 10.0 GPa.

We also observe that *P4/ncc* structure becomes more stable than the *R$\bar{3}$m* structure at 5.0 GPa. The phase transition pressure is determined to be 4.62 GPa at 0 K. The phase transition pressure for *P4/ncc* to *Im$\bar{3}$m* is 8.46 GPa at 0 K. However, the *Cmce* structure is more stable than *P4/ncc* or *Im$\bar{3}$m* at both pressures.

The *Cmce* structure is most stable at 5.0 GPa to 10.0 GPa. The *Pbcm* → *Cmce* phase transition pressure is 4.91 GPa at 0 K. The *P4/ncc* model structure is not a stable phase at this pressure range.

### 3.1.5 Crystal structures of Bi at 11.0 GPa

At 11.0 GPa, the phase transition to the *Im$\bar{3}$m* structure happens and the *Im$\bar{3}$m* structure is 1.6 meV/atom more stable than the *Cmce* structure. The densities of the *Cmce* and *Im$\bar{3}$m* structures are 12.54 and 12.76 g/cm$^3$, respectively. The density change is about 1.7%. The phase transition pressure for *Cmce* → *Im$\bar{3}$m* is determined to be 10.57 GPa.

Meanwhile, the *P4/ncc* structure is still 8.5 meV/atom less stable than the *Im$\bar{3}$m* structure. The phase transition pressure from *Cmce* to *Im$\bar{3}$m* structure is about 10.41 GPa. Previous calculations by Häussermann[35] reported that the transition pressure for *P4/ncc* → *Im$\bar{3}$m* was at 15.5 GPa. However, in our calculation, the inclusion of SOI shifted the transition pressure to 8.46 GPa. We also calculated the transition pressure for *P4/ncc* → *Im$\bar{3}$m* without the SOI, and the transition pressure is calculated to be 13.32 GPa. This result also agrees well with previous calculation by Dong *et al*[21], who reported the phase transition for *P4/ncc* → *Im$\bar{3}$m* structure at 13.4 GPa. Our calculation shows that SOI is very important for the phase transition of bismuth.

In summary, for high pressure phase transition of Bi at low temperature, four phases are predicted by our structure search method, namely *R$\bar{3}$m*, *Pbcm*, *Cmce* and *Im$\bar{3}$m*. The predicted phase transition pressures at 0 K are 3.29 GPa (*R$\bar{3}$m* → *Pbcm*), 4.91 GPa (*Pbcm* → *Cmce*) and 10.57 GPa (*Cmce* → *Im$\bar{3}$m*). The Bi-III modeled with *P4/ncc* structure, is less stable.

**3.2 Phase transition of bismuth at finite temperature**

Since the phase transition of bismuth is sensitive to temperature, we tried to take the finite temperature effect into consideration by using QHA method. We have successfully applied the QHA method to analyze the stability of different zirconium hydrides at finite temperature [29]. In our calculation, we found that at higher temperature the phase transition pressure for *R$\bar{3}$m* → *Pbcm*, *Pbcm* → *Cmce* and *Cmce* → *Im$\bar{3}$m* decreases. The simulated phase diagram is plotted in Figure 2. The phase transition pressures and temperatures are listed in Table 2.

At 2.5 GPa, the *R$\bar{3}$m* structure is most stable at 0 K, but as the temperature increases, the phase transition from *R$\bar{3}$m* to *Pbcm* happens at 305 K. At 2.75 GPa and 3.00 GPa, the phase transition temperatures are reduced to 217 K and 77 K, respectively.

For the phase transition *Pbcm* → *Cmce*, the transition pressure at 0 K is 4.91 GPa. However, at 3.25 GPa, 3.50 GPa, 3.75 GPa, 4.00 GPa and 4.50 GPa, the phase transition temperature are 497 K, 416 K, 342 K, 271 K and 116 K, respectively. The same phenomenon is also observed for the phase transition of *Cmce* → *Im$\bar{3}$m*. The phase transition temperature at 7.0 GPa, 8.0 GPa, 9.0 GPa and 10.0 GPa are 530 K, 398 K, 249 K and 66 K, respectively. Our calculation results agree with the trend of experimental phase diagram.

However, the Bi-III phase modeled by the *P4/ncc* structure is not a stable phase in our calculation. As a comparison, we also calculated the phase transition temperature for *R$\bar{3}$m* → *P4/ncc* and *P4/ncc* → *Im$\bar{3}$m*. For *R$\bar{3}$m* → *P4/ncc* phase transition, the phase transition temperature at 3.0 GPa and 3.5 GPa are 485 K and 325 K, respectively. For *P4/ncc* → *Im$\bar{3}$m* phase transition, the phase transition pressure at 6 GPa, 7 GPa are 212 K and 130 K, respectively.

**4 Conclusions**

We have performed structure search for Bismuth at different pressure. From our structure prediction calculation, we found a new phase of bismuth at low temperature with symmetry of *Pbcm*. There are four stable phases for bismuth at low temperature: *R$\bar{3}$m, Pbcm, Cmce* and *Im$\bar{3}$m*.

The phase transition pressures at 0 K are 3.29 GPa ($R\bar{3}m \rightarrow Pbcm$), 4.91 GPa ($Pbcm \rightarrow Cmce$) and 10.57GPa ($Cmce \rightarrow Im\bar{3}m$). When the temperature effect is considered by QHA method, we found that the phase transition pressure decreases as the temperature increase. The phase diagram of bismuth from first principles calculations is plotted. The *P4/ncc* structure, which is often adopted as model structure for Bi-III phase, is not a good model structure for studying the phase transition of bismuth. The SOI effect is very important in the calculation of bismuth and cannot be omitted.

**Acknowledge**

This work was supported by the Science Challenge Project (Grant No. TZ2018002，TZ2016001) and the National Science Foundation of China (Grant No. 51701015, 91730302, 11050504, 11501039). The authors also acknowledge computing resources from Special Program for Applied Research on Super Computation of the NSFC-Guangdong Joint Fund (the second phase) under Grant No. U1501501 and the National Key Research and Development Program of China under Grant No. 2016YFB0201204.

**References**

[1] Brugger R M, Bennion R B, Worlton T G. The crystal structure of bismuth-II at 26 kbar. Phys Lett A, 1967, 24: 714-717.
[2] Chen J H, Iwasaki H, Ktkegawa T. Crystal structure of the high pressure phases of bismuth bi iii and bi iii' by high energy synchrotron x-ray diffraction High Pressure Research, 1996, 15: 143-158.
[3] Mcmahon M I, Degtyareva O, Nelmes R J. Ba-IV-Type Incommensurate Crystal Structure in Group-V Metals Phys Rev Lett, 2000, 85: 4896-4899.
[4] Bridgman P W. Polymorphism, Principally of the Elements, up to 50,000 kg/$cm^2$. Phys Rev, 1935, 48: 893-906.
[5] Chen H-Y, Xiang S-K, Yan X-Z, et al. Phase transition of solid bismuth under high pressure. Chin Phys B, 2016, 25(10): 108103.
[6] Aoki K, Fujiwara S, Kusakabe M. Stability of the bcc Structure of Bismuth at High Pressure J Phys Soc Jpn, 1982, 51: 3826-3830.
[7] Liu L, Song H X, Geng H Y, et al. Compressive behaviors of bcc bismuth up to 55GPa Phys Status Solidi B, 2013, 250: 1398-1403.
[8] Akahama Y, Kawamura H, Singh a K. Equation of state of bismuth to 222 GPa and comparison of gold and platinum pressure scales to 145 GPa. J Appl Phys, 2002, 92: 5892-5897.
[9] Mukherjee D, Sahoo B D, Joshi K D, et al. On equation of state, elastic, and lattice dynamic stability of bcc bismuth under high pressure: Ab-initio calculations. J Appl Phys, 2014, 115(5): 053702.
[10] Chen J H, Iwasaki H, Kikegawa T. STRUCTURAL STUDY OF THE HIGH-PRESSURE-HIGH-TEMPERATURE PHASE OF BISMUTH USING HIGH ENERGY SYNCHROTRONSRADIATION. Journal of Physics and Chemistry of Solids, 1997, 58: 247 - 255.
[11] Degtyareva V F. Crystal structure of a high-pressure phase in Bi-based alloys related to Si VI. Phys Rev B, 2000, 62: 9-12.
[12] Chaimayo W, Lundegaard L F, Loa I S, G. W., et al. High-pressure, high-temperature

single-crystal study of Bi-IV High Pressure Research, 2012, 32: 442-449.

[13] Homan C G. Phase diagram of Bi up to 140 kbars. J Phys Chem Solids, 1975, 36: 1249 - 1254.

[14] Glass C W, Oganov a R, Hansen N. USPEX: Evolutionary crystal structure prediction. Comput Phys Commun, 2006, 175(11-12): 713 - 720.

[15] Wang Y, Lv J, Zhu L, et al. CALYPSO: A method for crystal structure prediction. Comput Phys Commun, 2012, 183(10): 2063 - 2070.

[16] Shang C, Liu Z-P. Stochastic Surface Walking Method for Structure Prediction and Pathway Searching. J Chem Theory Comput, 2013, 9(0): 1838.

[17] Wang W, Liang Z, Han X, et al. Mechanical and thermodynamic properties of ZrO2 under high-pressure phase transition: A first-principles study. J Alloys Compd, 2015, 622(0): 504 - 512.

[18] Shang C, Zhao W-N, Liu Z-P. Searching for new TiO2 crystal phases with better photoactivity. J Phys: Condens Matter, 2015, 27(13): 134203.

[19] Zhong X, Wang H, Zhang J, et al. Tellurium Hydrides at High Pressures: High-Temperature Superconductors. Phys Rev Lett, 2016, 116: 057002.

[20] Shu Y, Hu W, Liu Z, et al. Coexistence of multiple metastable polytypes in rhombohedral bismuth. Scientific Reports, 2016, 6: 20337.

[21] Dong X, Fan C-Z. First principle study of Bismuth under high pressure. Journal of Yanshan University, 2014, 38: 497.

[22] JASMIN，http://www.caep-scns.ac.cn/JASMIN.php.

[23] Mo Z, Zhang A, Cao X, et al. JASMIN: a parallel software infrastructure for scientific computing. Frontiers of Computer Science in China, 2010, 4(4): 480-488.

[24] Fang J, Gao X, Song H, et al. On the existence of the optimal order for wavefunction extrapolation in Born-Oppenheimer molecular dynamics. J Chem Phys, 2016, 144(24): 244103.

[25] Gao X, Mo Z, Fang J, et al. Parallel 3-dim fast Fourier transforms with load balancing of the plane waves. Comput Phys Commun, 2017, 211: 54-60.

[26] Wales D J, Doye J P K. Global optimization by basin-hopping and the lowest energy structures of Lennard-Jones clusters containing up to 110 atoms. J Phys Chem A, 1997, 101(28): 5111-5116.

[27] Zhao Y, Chen X, Li J. TGMin: A global-minimum structure search program based on a constrained basin-hopping algorithm. Nano Research, 2017, 10(10): 3407-3420.

[28] Chen X, Zhao Y-F, Wang L-S, et al. Recent Progresses of Global Minimum Searches of Nanoclusters with a Constrained Basin-Hopping Algorithm in the TGMin Program. Computational and Theoretical Chemistry, 2017, 1107: 57-65.

[29] Zhu X, Lin D-Y, Fang J, et al. Structure and thermodynamic properties of zirconium hydrides by structure search method and first principles calculations. Computational Materials Science, 2018, 150: 77-85.

[30] Kresse G, Furthmüller J. Efficiency of ab-initio total energy calculations for metals and semiconductors using a plane-wave basis set. Computational Materials Science, 1996, 6(1): 15-50.

[31] Perdew J P, Burke K, Ernzerhof M. Generalized gradient approximation made simple. Phys Rev Lett, 1996, 77(18): 3865-3868.

[32] Perdew J P, Burke K, Ernzerhof M. Generalized Gradient Approximation Made Simple [Phys. Rev. Lett. 77, 3865 (1996)]. Phys Rev Lett, 1997, 78(7): 1396-1396.

[33] Blöchl P E. Projector augmented-wave method. Phys Rev B, 1994, 50: 17953-17979.

[34] Arnaud B, Leb\`Egue S, Raffy G. Anisotropic thermal expansion of bismuth from first principles. Phys Rev B, 2016, 93: 094106.

[35] Häussermann U, Söderberg K, Norrestam R. Comparative Study of the High-Pressure Behavior of As, Sb, and Bi. J Am Chem Soc, 2002, 124(51): 15359-15367.

[36] Yoneda A, Endo S. Phase transitions in barium and bismuth under high pressure. J Appl Phys, 1980, 51(6): 3216-3221.

[37] Compy E M. Phase Diagram of Bismuth at Low Temperatures. J Appl Phys, 1970, 41(5): 2014-2018.

[38] Degtyareva O, Mcmahon M I, Nelmes R J. High-pressure structural studies of group-15 elements. High Pressure Research, 2004, 24: 319-356

[39] Compy E M. Phase Diagram of Bismuth at Low Temperatures J Appl Phys, 1970, 41: 2014-2018.

Figrue Captions

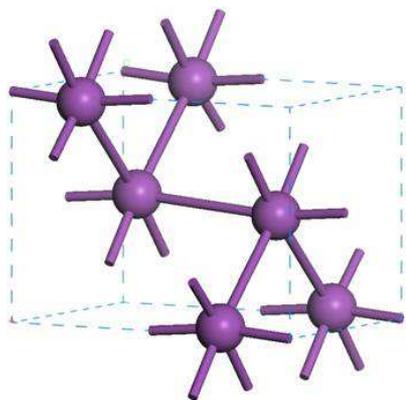
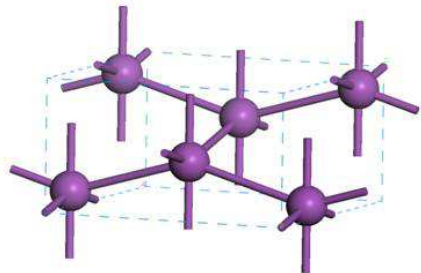
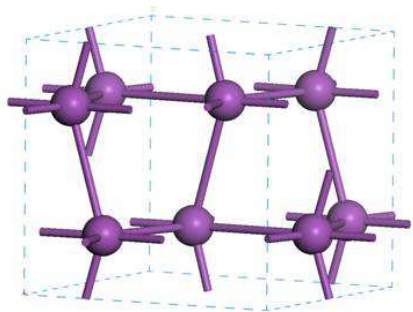

Figure 1 Structure of *Imma*(S1), *Imma*(S2) and *Pbcm* bismuth at 3 Gpa.

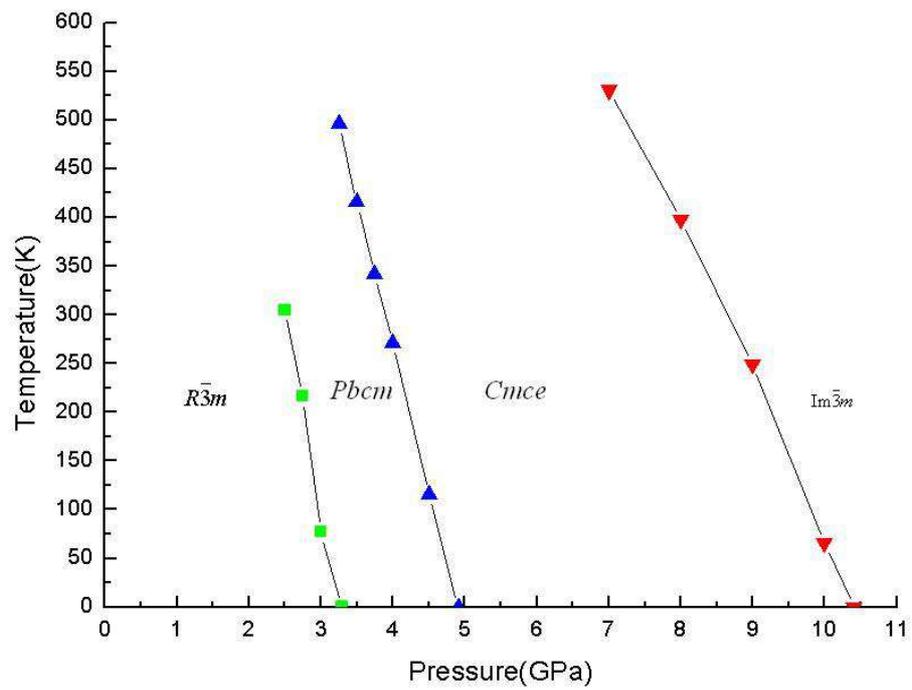

Figure 2 Pressure-Temperature phase diagram of bismuth simulated from first principles calculations.

Table captions

Table 1 Lattice constants, densities and enthalpy of different Bi crystals at 0~11 GPa

| Pressure(GPa) | Symmetry | #Atoms | Lattice Constants (Å and °) | Density (g/cm$^3$) | Enthalpy (meV/atom) |
|---|---|---|---|---|---|
| 0 | $R\bar{3}m$ | 6 | $a = b = 4.579, c = 12.184$ | 9.406 | 0.0 |
|  | Imma (S1) | 4 | $a = 6.520, b = 4.558, c = 4.955$ | 9.428 | 1.9 |
|  | $I4/mmm$ | 4 | $a = b = 4.609, c = 6.902$ | 9.469 | 5.8 |
|  | Pbcm | 8 | $a = 6.354, b = 6.738, c = 6.354$ | 10.207 | 48.3 |
|  | Imma (S2) | 4 | $a = 6.384, b = 6.327, c = 3.368$ | 10.206 | 50.9 |
| 3.0 | $R\bar{3}m$ | 6 | $a = b = 4.514, c = 11.534$ | 10.231 | 0.0 |
|  | Imma (S1) | 4 | $a = 4.505, b = 6.426, c = 4.689$ | 10.244 | 1.7 |
|  | Pbcm | 8 | $a = 6.281, b = 6.818, c = 6.195$ | 10.781 | 3.7 |
|  | Cmce | 16 | $a = 16.204, b = 6.638, c = 6.603$ | 11.308 | 21.5 |
|  | $P4/ncc$ | 32 | $a = b = 8.716, c = 12.983$ | 11.259 | 22.9 |
| 5.0 | Cmce | 16 | $a = 11.062, b = 6.576, c = 6.547$ | 11.659 | 0.0 |
|  | Pbcm | 8 | $a = 6.310, b = 6.485, c = 6.065$ | 11.186 | 0.8 |
|  | $P4/ncc$ | 32 | $a = b = 8.628, c = 12.840$ | 11.619 | 3.1 |
|  | $I4_1/amd$ | 4 | $a = b = 6.172, c = 3.280$ | 11.108 | 4.3 |
|  | $Im\bar{3}m$ | 2 | $a = b = c = 3.876$ | 11.917 | 15.9 |
|  | $R\bar{3}m$ | 6 | $a = b = 4.476, c = 11.281$ | 10.639 | 19.9 |
| 11.0 | $Im\bar{3}m$ | 2 | $a = b = c = 3.789$ | 12.764 | 0.0 |
|  | Cmce | 16 | $a = 10.746, b = 6.43, c = 6.409$ | 12.538 | 1.6 |
|  | $P4/ncc$ | 32 | $a = b = 8.424, c = 12.518$ | 12.501 | 8.5 |

Table 2 Wyckoff positions and coordinates of the most stable structures at 3.0 GPa

| Space group | Wyckoff position | Coordinates |
|---|---|---|
| $R\bar{3}m$(#166) | 6 c | (0.00000, 0.00000, 0.23817) |
| $Imma$(S1)(#74) | 4 e | (0.50000, 0.25000, 0.52887) |
| $Imma$(S2)(#74) | 4 e | (-0.50000, 0.75000, 0.38529) |
| $Pbcm$(#57) | 4 d | (0.99169, 0.31788, 0.25000) |
|  | 4 d | (0.49243, 0.43256, 0.25000) |
| $Cmce$(#64) | 8 d | (0.50000, 0.82348, 0.32918) |
|  | 8 f | (0.70847, 0.50000, 0.50000) |
| $P4/ncc$(#130) | 16 g | (0.34079, 0.35047, 0.91705) |
|  | 8 f | (0.84377, 0.15623, 0.25000) |
|  | 4 c | (0.00000, 0.50000, 0.95589) |
|  | 4 c | (0.00000, 0.50000, 0.19358) |

Table 3 Phase transition pressures and temperatures for Bismuth from first principles calculations

| phase transition | Pressure(GPa) | Temperature(K) |
| --- | --- | --- |
| $R\bar{3}m \rightarrow Pbcm$ | 2.5 | 304.7 |
| | 2.75 | 216.8 |
| | 3 | 77.1 |
| | 3.29 | 0.0 |
| Pbcm→Cmce | 3.25 | 496.5 |
| | 3.5 | 415.9 |
| | 3.75 | 342 |
| | 4.0 | 271.3 |
| | 4.5 | 115.8 |
| | 4.91 | 0 |
| Cmce → $Im\bar{3}m$ | 7.0 | 530.4 |
| | 8.0 | 397.9 |
| | 9.0 | 249.3 |
| | 10.0 | 65.7 |
| | 10.41 | 0 |